\documentclass{hbarticle}
\usepackage{graphicx}
\usepackage{theorem}
\usepackage{amsfonts}
\usepackage{amssymb}
\usepackage{subfiles}
\usepackage{xcolor}
\usepackage{fancyhdr}

\normalsize
\title{Deep Bellman Hedging\\{\small Revision 2.03 April 2024}}
\date{First version London June 30th, 2022}
\author{Hans Buehler\footnote{Technical University M\"unchen}, Phillip Murray\footnote{JP Morgan London, Imperial College London}, Ben Wood\footnote{JP Morgan London}}

\makeindex

\begin{document}

    \maketitle
    
    \begin{abstract}
        We discuss an
        actor-critic-type reinforcement learning algorithm for solving the 
        problem of hedging a portfolio of financial instruments
        such as securities and over-the-counter derivatives
        using purely historic data.  

    Essentially we present a practical implementation for 
    solving a problem of the structure
    \[
            V^*\big(\,
                \mbox{current state}
            \,\big)
            \supstack!=
            \sup_{\mbox{action $\ma$}}:\
                 U\!\left[\ \mbox{discount factor $\cdot$} \
                       V^*\big(\,
               \mbox{future state}(\ma)
            \,\big) + \mbox{rewards}(\ma)
                \ \right]  
     \]
     for the optimal value function~$V^*$
     where $U$ represents a risk-adjusted return metric.

        The key characteristics of our approach are: the ability to hedge with derivatives such as forwards, swaps, futures, options;
    incorporation of trading frictions such as trading cost and liquidity constraints;
    applicability to any reasonable portfolio of financial instruments; realistic, continuous state and action spaces;
    and formal risk-adjusted return objectives. We do not impose a boundary condition on some future maturity.

        Most importantly,
        the trained model provides an optimal hedge for arbitrary initial portfolios and
        market states without the need for re-training.

        We also prove existence of finite solutions to our Bellman equation, 
        and show the relation 
        to our vanilla Deep Hedging approach~\cite{DH}
    \end{abstract}

\noindent
    \emph{The views expressed on this note are the author’s personal views and should not be attributed to any other person, including that of their employers.}

    \tableofcontents

    \section{Introduction}
    
    This note discusses a model-free, data-driven
    method of managing a portfolio of financial instruments such
    as stock, FX, securities and derivatives
    with \emph{reinforcement learning} "AI" methods.
    It is a dynamic programming ``Bellman" version of our
    Deep Hedging approach~\cite{DH}.

    \subsubsection*{Quant Finance 2.0}
    
    The motivation for the work presented in this article -- and of the Deep Hedging framework in 
    general -- is to build financial risk management models which ``learn" to trade from historic
    data and experiences. Today, portfolios of derivatives, securities and other instruments
    are managed using the traditional quantitative finance engineering paradigm borne out of the seminal work by
    Black, Scholes~\& Merton. However, practical experience on any trading desk is that 
    such models do not perform sufficiently well to be automated directly. To start with, they
    do not take into account trading frictions such as cost and liquidity constraints. Even beyond
    that they suffer from the underlying engineering approach which prioritizes a focus on interpolating
    hedging instruments such as forwards, options, swaps over realistic market dynamics.\footnote{See also the discussion in~\cite{NEARMART}.}
    It is an indication of the state of affairs that standard text books on financial engineering
    in quantitative finance do not discuss real data out-of-sample performance of the models proposed.

    As a result,
    the prices and risk management signals (``greeks") are overly simplistic and do not capture
    important real-life dynamics. A~typical trader will therefore need to adjust
    the prices and
    trading signals
    from such as standard models using their own heuristics.

    \subsubsection*{From Deep Hedging to Deep Bellman Hedging}

    Our \emph{Deep Hedging} framework~\cite{DH}, \cite{MULTIASSET}, 
    \cite{NEARMART_ARXIV}, \cite{NEARMART} takes a different approach and focuses on robust performance
    under real-life dynamics, enabed by the use of modern machine learning techniques. 
    Its key advantages  above
     other proposed methods for hedging a portfolio of financial instruments are:
    \begin{itemize}
        \item  hedging with any number of reasonable derivatives such as forwards, swaps, futures, options;
        
        \item incorporation of trading frictions such as trading cost and liquidity or
        risk constraints;
        
        \item applicability to any reasonable portfolio of financial instruments, not
        just portfolios of primary assets;
        
        \item realistic, continuous state and action spaces; and
        
        \item 
             provision for formal risk-adjusted return objectives.

    \end{itemize}
    
   Our original approach~\cite{DH}
    solved this problem for a given initial portfolio and market state.
    That means it needs to be re-trained, say, daily to reflect changes in our
    trading universe or the market. 
    The method 
    proposed here, on the other hand, attempts to solve the optimal hedging problem over an
    for any portfolio and market state which is close to past experiences.
    It does not impose a boundary restriction for some terminal maturity.

    Our main Bellman equation~\eqref{bell_U} for the optimal value function~$V^*$ has the structural form
    \[
            V^*\big(\,
                \mbox{current state}
            \,\big)
            \supstack!=
            \sup_{\mbox{action $\ma$}}:\
                 U\!\left[\ \mbox{discount factor $\cdot$} \
                       V^*\big(\,
               \mbox{future state}(\ma)
            \,\big) + \mbox{rewards}(\ma)
                \ \right]  
     \]
     where $U$ represents a risk-adjusted return metric.

    The work presented here is an extension of
    our patent application~\cite{PATENT_DH}.
    The main contribution is to provide
    a numerical implementation method for the practical problem of being able to represent
    arbitrary portfolios of derivatives as states using purely historic data. 
    This means it does not require the development of a market simulator, c.f.~\cite{MULTIASSET}.
    
        The current article is the closest attempting to mimicking a trader's real life behaviour
    in that here we will give an AI the same historic ``experience" a real trader would have.
    Of course, our model will still be limited by the coverage of historic scenarios used
    to train it. Hence, human oversight is still required to cater for abrupt changes in
    market scenarios or starkly adverse risk scenarios.
    
    As a fundamental contribution we also clarify conditions
    under which the corresponding Bellman equations are well-posed and
    admit unique finite solutions.

The website \texttt{http://deep-hedging.com} gives an overview over available material on the
wider topic.

    \subsubsection*{Related Works}
    
    There a few related works concerning the use of machine learning methods for 
    managing portfolios of financial instruments which include derivatives, starting with our
    own~\cite{DH}. There, we solved the optimal trading problem for a fixed initial portfolio with a given terminal maturity
    and a fixed initial market state using \emph{periodic policy search}, a method akin to ``American Monte Carlo".

    In~\cite{RITTER_VOL} the authors discuss the use of Bellman methods for this task, namely  using DQN
    and a number of similar methods. They also use risk-adjusted returns in the form of a mean-variance
    objective. However, in their work the state and action spaces are 
    finite which is not realistic in practise. Moreover, their parametrization of the derivative
    portfolio is limited to single vanilla options. They also do not cover derivatives as hedging instruments. The method
    relies on a
    fixed terminal maturity.
    
    In~\cite{IGOR_DQN} the authors also develop a discrete state approach, 
    where the problem is solved for each derivative position separately. The authors focus in
    their first work on vanilla options and minimize the terminal variance of the delta-hedged position.
    The maturity of the problem is fixed.
    In their later~\cite{IGOR_DQN2} the authors present methods to smooth the state space.
    In neither account are derivatives as hedging instruments supported.
    
    A~forthcoming ICML contribution~\cite{DHBLT} we solve the Bellman equation
    associated with the Deep Hedging problem for a fixed maturity, using continuous states
    and the entropy as risk-adjusted return.
    
    There is a larger literature on the application of AI methods for managing portfolio risks in the context
    of perpetual primary assets such as stock and FX portfolios whose 
    distributions might reasonably be approximated
    by Gaussian variables. See the summary~\cite{RITTER_SUMMARY} for an overview, where they also cover
    the related topic of trade execution with AI methods.

    Underlying our work is the use of \emph{dynamic risk measures}, a topic
    with a wide literature. We refer the interested reader to~\cite{CONDCRM} and~\cite{PENNER}
    among many others.

    \section{Deep Bellman Hedging}
    
    In this note we will use a notation much more similar to standard
    reinforcement learning literature, chiefly~\cite{SUTTON}. That means
    in particular that we will formulate our approach essentially 
    as a continuous state Markov Decision Process (MDP) problem. We will make a decision
    from some point in time to another. That would typically be intraday or from day to day.
    To simplify our discussion we will assume we are making a decision ``today"
    and then again ``tomorrow". Variables which are valid tomorrow
    will be indicated by a~$'$. We will strive to use bold letters for
    vectors. A~product of two vectors is element wise, while ``$\cdot$"
    represents the dot product. We will use small letters
    for instances of data, and capital letters for random variables.\\
    
    We denote by~$\mm$ the \textbf{market state} today. The market contains all information available to us today such as
    current market prices, time, past prices, bid/asks, social media feeds and the like. The set of all market states is
    denoted by~$\calM\subset \R^N$.
    All quantities observed today are a function of the market state.\footnote{
    Mathematically, we say that~$\mm$ generates today's $\sig$-algebra}
    Past market states are also known today.\footnote{This means
    the sequence of market states generates a filtration.
    } 
	The market tomorrow is a random variable $\mM'$
	whose distribution is assumed to  only depend on~$\mm$,
	and not on our trading activity.\footnote{
		See the lecture notes~\cite{LECTURE3} for an example
		of incorporating market impact.
	}
    In terms of notation, think $\mm\equiv \mm_t$ and 
    $\mM'\equiv \mM_{t+1}$.
        The expectation operator of a function~$f$ of $\mM'$ conditional
    on $\mm$ is written as~$\E[f(\mM')|m]:=\int f(\mm') \P[d\mm'|\mm]$.
    
	We will trade financial instruments such as securities, OTC derivatives
	or currencies. We will loosely refer to them as ``derivatives"
	as the most general term, even if we explicitly include primary
	asset such as stocks and currencies. We use~$\calX$ to refer to the
	 space of these	instruments. 
	For~$x\in\calX$ we denote by $r(x,\mm)\in\R$ the  cashflows arising from holding~$x$ today, aggregated
	into our accounting currency.\footnote{This implies
    implies that spot-FX transactions
    are frictionless.}
    Cashflows here cover everything from expiry settlements, coupons, dividends, to payments
    arising from borrowing or lending an asset.
    For a vector $\mx$ of instruments we use $
    \mr(\mx;\mm)$ to denote the vector of their cashflows.

    An~instrument changes with the passage of time: an instrument~$x\in\calX$ today
    becomes $x'\in\calX$ tomorrow, representing only cashflows from tomorrow onwards.
    If the expiry of the instrument is today, then~$x'=0$.
    
    Every instrument~$x$ we may trade has a \textbf{book value} in our accounting currency
    which we denote by $B(x,\mm)$. 
     The book value of a financial instrument is its official 
     mark-to-market, computed
     using the prevailing market data~$\mm$. This could be a simple closing price, a weighted
     mid-price, or the result of running a
     more complex standard derivative model.
     Following our notation $B(x',\mM')$ denotes the book value of the instrument tomorrow.
     We use $\mB(\mx,\mm)$ for the vector
    of book values if~$\mx$ is a vector of instruments. We like
    to stress that contrary to our other work~\cite{NEARMART} here 
    the book value is with respect  only to
     today's and future
    cashflows, not past cashflows.

    In order to take into account the
    value of money across time, we will also assume are given a bank
    account -- usually called the \emph{numeraire} - which charges the 
    same overnight interest rate for
    credits and deposits. 
    The respective one-day discount factor from tomorrow to today
    is denoted by~$\beta(\mm)$ 
    and we will assume that there is some
    $\beta^*$ such that $\beta(\mm)\leq\beta^*<1$.
    Contrary to~\cite{DH} we do not assume that 
     cashflows are 
    discounted using the numeraire.
    
    The discounted profit-and-loss (P\&L) for a given instrument~$x\in\calX$
    is the random variable
    \eq{dB}
        dB(x,\mm,\mM') := \underbrace{ \beta(\mm)\, B(x',\mM') - B(x,\mm) }_{\mbox{Change in book value}} + \underbrace{ r(x,\mm) }_{\mbox{Cashflows}} \ .
    \eqend
    If $\mx\in\calX^n$ is a vector, then $d\mB(\mx,\mm,\mM')$ denotes the vector of P\&Ls.
    
	\subsubsection*{Trading}

    A~trader is in charge of a \textbf{portfolio} 
    $z\in\calX$ -- also called ``book" -- of financial
    instruments such as currencies, securities and over-the-counter (OTC) derivatives.
    We call the combined~$\ms:=(z,\mm)$ our \textbf{state} today
    which takes values in $\ms\in\calS:=\calX\times \calM$.
    We will switch in our notation between writing functions in both variables
    $(z,\mm)$ and only in~$(\ms)$ depending on context. 

    In order to risk manage her portfolio, the trader has access to~$n$  
    liquid hedging instruments $\mh\equiv \mh(\mm) \equiv \mh(\ms) \in\calX^n$ in each 
    time step.
    These are any liquid instruments such 
    such as forwards, options, swaps etc. 
    Across different market states they will usually not be the contractually same fixed-strike fixed-maturity instruments:
    instead, they will
    usually be defined relative the prevailing market 
    in terms of time-to-maturities
    and strikes relative to at-the-money. See~\cite{DH}
    for details.
    
     The \textbf{action} of buying\footnote{Selling amounts
     to purchasing a negative quantity.}~$\ma\in\R^n$ 
     units of our hedging instruments
     will incur transaction cost $c(\ma;z,\mm)$ on top of the book value.
     Transaction 
     cost as function of~$\ma$ is assumed to be normalized to $c(0;\ms)=0$, non-negative, and convex.\footnote{Convexity excludes fixed transaction cost.} 
     The convex set of admissible actions is given 
     as~$\calA(z,\mm):=\{\ma\in\R^n: c(\ma;z,\mm) < \infty\}$. 
     Making cost dependent on both the current portfolio
     and the market  allows modelling
     trading restrictions based on our current position such as short-sell constraints,
     or restrictions based on risk exposure.    

     Given $z$ and an action $\ma$ today the new portfolio tomorrow is $z'+\ma \cdot \mh' \in \calX$.\\

     A~\textbf{trading policy} $\pi$ is a function
     $\pi$ 
      which determines the next action based on our current state,
     i.e.~simply $\ma = \pi(z,\mm)$. We use $\calP:=\{ \pi: \calX\times \calM \rightarrow \calA(z,\mm) \}$ to refer to the convex set of
     all admissible policies.

    A~trader will usually manage her book by referring to the change in book values
    plus any other cashflows, most notably cashflows and the cost of hedging.
    The associated \textbf{reward}  for taking an action~$\ma$ per time step is given as
     \eq{rewards}
        R(\ma;z,\mm, \mM') := \underbrace{ dB(z,\mm,\mM') }_{
                    \begin{array}{c}
                        \mbox{Portfolio}\\
                        \mbox{P\&L}\end{array}
                    } + 
                    \underbrace{ \ma \cdot  d\mB(\mh,\mm,\mM') }_{
                        \begin{array}{c}
                        \mbox{New Hedge}\\
                        \mbox{P\&L}\end{array}
                    }
                    -
                    \underbrace{ c(\ma;z,\mm) }_{
                        \begin{array}{c}
                        \mbox{Trading}\\
                        \mbox{Cost}\end{array}
                        }
                    \ .
     \eqend
     For the rewards of a policy $\pi$ we will use the convention
     \[
        R(\pi;z,\mm, \mM') := R(\pi(z,\mm);\,z,\mm, \mM') \ .
     \]
     \removeblock{
     The new joint portfolio tomorrow is given by
     \eq{pftomorrow}
            z_\ma' := z' + \ma \cdot \mh' \ .
     \eqend
     The new state tomorrow is a random variable depending on our action which we write as
     \[
        \mS'_\ma := (z_\ma',\mM') \ .
     \]
    }
    
     \subsection{The Bellman Equation for Monetary Utilities}
    
    Standard reinforcement learning as discussed for example
    in~\cite{SUTTON} usually aims to maximize
    the discounted expected future rewards of running a given policy.
    Essentially, the optimal value function $V^*$ is stipulated to satisfy
    a Bellman equation of the form
    \[
        V^*\big(\,
            z;\,\mm
        \,\big)
         \supstack!= 
        \sup_{\ma\in \calA(z,\mm)}: \
           \E\!\left[\  \beta(\mm) \ 
                   V^*\big(\,
            z' +\ma\cdot \mh';\,\mM'
        \,\big) + R(\ma;z,\mm, \mM')
            \,\big|\,\mm
            \ \right]  \ .
    \]
    Instead of using the expectation 
   it is more natural in finance to 
    choose an operator~$U$ which takes into account risk aversion: 
   this 
    roughly means that if two events have the same expected outcome, then we prefer the
    one with the lower uncertainty. 
    
    \begin{definition}
    The \emph{Deep Bellman Hedging} problem is finding a \textbf{value function} $V^*$ which satisfies
        \begin{equation}\label{eq:bell_U}
        \left\{
        	\begin{array}{lcl}
            V^*\big(\,
                z;\,\mm
            \,\big)
            & \supstack!= &
            (TV^*)(z,\mm) \\
            &&
            \\
            (Tf)(z,\mm) & := &
            \sup_{\ma\in \calA(z,\mm)}:\
                 U\!\left[\ \beta(\mm) \
                       f\big(\,
                z' +\ma\cdot \mh';\,\mM'
            \,\big) + R(\ma;z,\mm, \mM')
                \,\big|\,\mm
                \ \right]  \ .
           \end{array}
        	\right.     
        \end{equation}
        (The action~$\ma$ in above $\sup$ operator is a function of the state $\ms=(z,\mm)$.)
    \end{definition}
    
   \removeblock{

    We may therefore write the operator as an optimization over policies
    $\pi$ as follows:
    \begin{equation}\label{eq:bell_U_pi}
        Tf(\,
            z;\,\mm
        \,\big)
        \equiv
        \sup_{\pi(z,\ms)\in \calA(z,\mm)}:
            U\!\left[\ \beta(\mm) \ 
                   f\big(\,
            z' +\pi(z,\mm)\cdot \mh';\,\mM'
        \,\big) + R(\ma;z,\mm, \mM')
                    \,\big|\,\mm
            \ \right]  \ .
    \end{equation}
    }
    
    We would like to stress that the value function here represents the ``excess value"
    of a portfolio over its book value: \emph{if $V^*$ were zero, that would mean the optimal risk-adjusted
    value for a portfolio were given as its book value.} Remark~\ref{rem:cashflow_rewards} makes this statement explicit.\\
    
    There are many different reasonable 
    risk-adjusted return metrics~$U$ used in finance, most notably mean-volatility, mean-variance
    and their downside versions. We refer the reader to the seminal~\cite{MV}.
    Mean-volatility in particular remains a popular choice for many practical applications.
    However, it is well known that mean-volatility, mean-variance
    and their downside variants are not monotone, 
    which means that even if $f(\ms)\geq g(\ms)$
    for all states~$\ms=(z,\mm)$ it is not guaranteed that $U[ f(\mS' )\geq U[ g(\mS' )]$, c.f.~\cite{SH}. The lack of monotonicity
    means that standard convergence proofs for the Bellman 
    equation do not apply; see section~\ref{sec:proofs}.

    We will here take a  more formal route and focus on 
    \emph{monetary utilities}.
    A~functional~$U$ is called a \textbf{monetary utility}  if it is normalized to $U(0)=0$, monotone increasing (more is better),\footnote{
        If $f\geq g$ then $U[f(\mS')]\geq U[g(\mS')]$.
        }
    concave (diversification works)\footnote{
        For $X=f(\mS'),Y=g(\mS')$ and $\alpha\in[0,1]$ we have $U[\alpha X + (1-\alpha)Y] \geq \alpha U[X]
        + (1-\alpha) U[Y]$.
    } and \emph{cash-invariant}. The latter 
    means that  $U[ f(\mS',\ms) + y(\ms) |\ms] = U[f(\mS',\ms)|\ms] + y(\ms)$ for any function $y(\ms)$. 
    The intuition behind this property is if we add a cash amount~$y$
    to our portfolio, then its monetary utility increases by this amount.\footnote{We have shown in~\cite{SH} that cash-invariance is equivalent
    to being able to write-off parts of our portfolio for the worst
    possible outcome.} An important implication of cash-invariance is that our optimal actions
    do not depend on our current wealth. We will also assume that our monetary utilities
    are \emph{risk-averse with respect to~$\P$} in the sense that $\E[ f(\mS') ]=U[\E[ f(\mS') ]] \geq U[ f(\mS') ]$.\footnote{
        We note that concavity of $U$ does \emph{not} imply risk aversion w.r.t.~$\P$ in general. As an example chose a measure~$\Q\not\approx \P$
        and set $U(f(\mS')) := \E_\Q[ f(\mS') ]$.
    }

    The negative
    of a monetary utility is  called a \emph{convex risk measure}, c.f.~\cite{FS}. See also~\cite{CONDCRM} on the topic of dynamic
    and time-consistent risk measures.
    
    As in~\cite{NEARMART} we will focus on monetary utilities given 
    as \emph{optimized certainty equivalents} (\textbf{OCE}) of a utility function,
    introduced by~\cite{BT}.
    
    \begin{definition}
        Let $u:\R\rightarrow\R$ be a $C^1$ \emph{utility function} which 
        means it is
        monotone increasing
        and concave. We  normalize it to~$u(0)=0$ and $u'(0)=1$.
        The respective \emph{OCE monetary utility} is then defined by
        \[
            U\!\left[\  f(\mS') \ \big|\ \ms\ \right]
            :=
            \sup_{y(\ms)\in\R} \E\!\left[\ u\left(\ f(\mS')
            + y(\ms)\, \right)\ \big|\ \ms\ \right] - y(\ms) 
        \]
     \end{definition}
        The function~$y$ in our definition will be modelled as a neural network.

        Examples of OCE utility functions are
		\begin{itemize}
			\item
			\textbf{Expectation} (risk-neutral):
			$
				u(x) :=  x
			$.			
			\item
			\textbf{Worst Case}:
			$
				u(x) :=  \inf x
			$.			
			\item
			\textbf{CVaR} or \textbf{Expected Short Fall}:
			$
			u(x) := (1+\lambda)\ \min\{0,X\}
			$.

			\item	
			\textbf{Entropy}:
			$
			    u(x) := (1-e^{-\lambda x})/\lambda
		 $
			in which case $U[ f(\mS')|\ms] = - \frac 1\lambda
			\E[ \exp(-\lambda f(\mS')) | \ms ]$. The entropy
			reduces to mean-variance if 
			the variables concerned are normal. It has many
			other desirable properties, but it also 
			penalizes losses rather harshly: an unhedgable short position
			in a Black\&Scholes stock has negative infinite entropy.

			\item
			\textbf{Truncated Entropy}: to avoid the harsh penalties for short
			positions imposed by the exponential utility we might instead use
			$
				u(x) := 
					( 1- e^{-\lambda x} ) /\lambda \1[ x > 0 ]
					+
					( x - \half \lambda x^2 ) \1[x<0]$.
			
			\item
			\textbf{Vicky}: the following functional was proposed in~\cite{VICKY}:
			$
				u(x) := \frac1\lambda \left( 1 + \lambda x - \sqrt{ 1 + \lambda^2 x^2 } \right) 
			$.
			
			\item
			\textbf{Normalized quadratic utility}:
			$
				u(x) := - \half \lambda (x-\mbox{$\frac1\lambda$})^2\1[x<\mbox{$\frac1\lambda$}] + \frac1{2 \lambda} 
			$.
 		\end{itemize}

	\noindent
 		We call a monetary utility \emph{coherent} 
		if $U[ n(\ms)\,f(\mS' ) | \ms ] = n(\ms) U[ f(\mS') | \ms ]$.
		An OCE monetary utility is coherent if $u(n\,x) = n\,u(nx)$.
		 Coherence is not a particularly natural property: it says
		that the risk-adjusted value of a position growths linearly with position
		size. Usually, we would assume that risk increases superlinearly.
		The practical relevance of this property for us is that if~$U$
		is coherent, then we can move the discount factor~$\beta$
		in and out of our monetary utility: $\beta(\ms)\, U[ f(\mS') | \ms] = 
		U[ \beta(\ms)\,f(\mS')|\ms]$.		
		
    	We say $U$ is 		
 		 \emph{time-consistent}
 		if iterative application lead to the same monetary utility in the sense that $U[ U[ f(\mS'') | \mS' ] | \ms ] = U( f(\mS'' | \ms )$.  The only
 		time-consistent OCE monetary utilities are the entropy and the
 		expectation, c.f.~\cite{KUPPERSCH}. 
		\\

		\noindent
	We may now present the first key result of this article: recall the
	definition of our rewards~\eqref{rewards}.
	We say that \emph{rewards are  finite} if
		\eq{rewards_finite}
			\sup_{\ma\in\calA(z,\mm)} \, U\!\left[\,R(\ma;\mz,\mm,\mM')\,\big|\,\mm\,\right] <\infty	\ .
			\eqend
	for all $z\in\calX,\mm\in\calM$.
		This can be achieved for example if~$\calA(z,\mm)$ is bounded.	
 		
 	\begin{theorem}\label{th:convergence}
		Assume rewards are finite and that $U$ is an OCE monetary utility.
		
		Then the Bellman
		equation~\eqref{bell_U}
		has a unique finite solution.
	\end{theorem}
	The proof can be found in section~\ref{sec:proof_convergence}. It relies
	on monotonicity
	and cash-invariance of the monetary utility, which means that while it
	does not apply to mean-variance, mean-volatility and related operators,
	it does apply to $U$ being VaR, e.g.~the percentile function
	$U(X) := \P[X \leq \cdot]^{-1}( 1-\alpha )$ where $\alpha\in[0,1)$ is a confidence interval. 
	This is a good alternative to mean-volatility as it has a similar interpretation,
	as long as VaR's lack of concavity is not a concern.

    \begin{remark}[Using only Cashflows as Rewards]\label{rem:cashflow_rewards}
    
        Our definition of our rewards~\eqref{rewards} as the full mark-to-market of
        the hedged portfolio is in so far unusual as the reward term contains 
        future variables, namely the book value of the hedged book tomorrow.
        
        A~more classic approach would be to let the rewards represent
        only actual cashflows, e.g.
        \eq{discrewards}
            \tilde R(\ma,z,\mm):= \underbrace{ r(z,\mm) }_{
                \begin{array}{c}
                    \mbox{Cashflows from}\\
                    \mbox{our portfolio}
                \end{array}
                    } 
            \underbrace{ - \ma \cdot \mB(\mh,\mm) - c(\ma,z,\mm) }_{
                \begin{array}{c}
                    \mbox{Proceeds from}\\
                    \mbox{trading~$\ma$}
                \end{array}
                } \ ,
        \eqend
    The numerical challenge with this formulation is that cashflows are
     relatively rare for most hedging instruments: if we trade a 1M vanilla
    option, then it only has one cashflow at maturity -- if it ends up in the money. 
    That means that learning the value of future cashflows is harder
    when we train with only daily cashflows. Hence, it is numerically
    more efficient to solve for the difference between the optimal value
    function and the book value of a portfolio.
    
    Theoretically, though, the two are equivalent:
    let $\tilde V^*(z,\mm) := V^*(z,\mm) + B(z,\mm)$. Then~$\tilde V^*$
    solves the Bellman equation
    \eq{alternative}
    	\tilde V^*(z,\mm) \supstack!= \sup_{\ma\in\calA(\ms)}:
                    \
                    U\!\left[ \ 
                    \beta(\mm)
                    \
                    \tilde V^*(\,z' +\ma\cdot \mh',\mM\,)
                    + \tilde R(\ma,z,\mm)
                    \ \big|\ \mm\ \right] \ .   
    \eqend

    \end{remark}

	\begin{remark}[Multiple time steps]\label{rem:multitime}

    It is straight forward to formally extend~\eqref{bell_U} to multiple
    time steps. Let $\mS^{(1)}:=\ms$ and $\mS^{(i+1)}=\mS^{(i)}{}'$.
   	We use the same numbering for other variables.
   	 Define
    the discount factors
    $\beta_i := \prod_{e=1}^i \beta(\mM^{(e-1)})$ and set
    \begin{equation}\label{eq:bell_U_multi}
        (T_n f)(z,\mm) :=
        \sup_{\pi}:\
             U\!\left[\ \beta_n\
                   f\big(\,
            z^{(n)} + \mA^{(n)}\cdot \mH^{(n)};\,\mM^{(n+1)}
        \,\big) + \sum_{i=1}^n 
			\beta_{i-1}\,
        R(\mA^{(i)},z^{(i)},\mM^{(i)},\mM^{(i+1)}) 
             \ \big|\ \mM^{(1)}=\mm
            \ \right]  
    \end{equation}
    where we used $
    	\mA^{(n)} := \pi(z^{(n)}, \mM^{(n)})$.
    	Our indexing scheme means that $T_1 = T$.

	It is straight forward to 
	amend the proof of theorem~\ref{th:convergence} to
	show that if any statistical arbitrage is finite,
	then the associated equation $T_n f = f$ also has a unique finite 
	solution~$V^*_n$.
	
	However, the resulting value functions are not usually consistent: for $n>1$ in general $V^*_n \not= T^n V^*$ unless
	$U$ is the expectation.\footnote{
		We show the claim for $n=2$. Let $R^{(n)}:=R(\mA^{(n)},\mS^{(n)},\mM^{(n+1)} )$.
		We use the state notation $\ms=(z,\mm)$.
		\eqary
			T^2f(\ms)
			& = & \sup_\pi: U\left[\, \beta(\mM^{(1)})\
			\left\{
				\sup_{\pi}
							\,U\left[\,
								\beta(\mM^{(2)})\
								f(  \cdots, \mM^{(3)} )
								+
								R^{(2)}
								\,\big|\,\mS^{(2)}\,
								\right ]
			\right\} + R^{(1)}
			\,\big|\,\ms\,
			\right]
			\\
			& \supstack{(*)}= & \sup_\pi: U\left[\, \beta(\mM^{(1)})\
			\left\{
							\,U\left[\,
								\beta(\mM^{(2)})\
								f( \cdots, \mM^{(3)} )
								+
								R^{(2)}
								\,\big|\,\mS^{(2)}\,
								\right ]
			\right\} + R^{(1)}
			\,\big|\,\ms\,
			\right]
			\\
			& \supstack{(**)}{\geq/=} & \sup_\pi: U\left[\, \
							\,U\left[\,
								\beta(\mM^{(1)}) \beta(\mM^{(2)})\
								f( \cdots, \mM^{(3)} )
								+
								R^{(2)}
								+ \beta(\mS^{(1)}) R^{(1)}
								\,\big|\,\mS^{(2)}\,
								\right]
			\,\big|\,\ms\,
			\right]
			\\
			& \supstack{(***)}{=} & \sup_\pi U\left[ \cdots |\,\ms\,\right]
			\ = \ T_2 f(\ms) \ .
		\eqaryend
		    The equation $(*)$ is true if $\pi$ is found over the entire set of measurable
		    functions, and if the state contains all previous state information. 
			Moreover, $(**)$ is an equality for the expectation 
			or any other coherent monetary utility. For all others
			convexity and $U(0)=0$ imply the stated inequality.
			The final equality~$(**)$ is only true of $U$ is time-consistent
			which means either the entropy or the expectation.
			\qed
	}

    \end{remark}

    \section{Numerical Implementation}\label{sec:iterativeDBH}
    
    We now present an algorithm which will iteratively approach an optimal solution of our Bellman equation~\eqref{bell_U}.
    This is an extension over the entropy case presented in~\cite{PATENT_DH}.
    
    Recall that $V^*(z,\mm)=0$ is the optimal solution if the book value of the
    portfolio~$z$
    is indeed its value function. We therefore use as initial guess $V^{(0)}(\,z,\mm\, ) := 0$
    for all portfolios and states~$\ms=(z,\mm)$.\footnote{
    	It is not a good idea to initialize a network with zero to achieve this as all gradients will look rather the same. Assume $\calN(\theta;x)$ is a neural network initialized by random
    	weights $\theta_0\in\R^W$. Then use
    	the \emph{Buehler-zero} network
    	$N(\theta;x) := \calN(\theta;x) - \eta
    	\calN(\theta_0;x)$ and learn $(\theta, \eta)$ where $\eta\in[0,1]$. 
    } 
    
    We then solve our problem iteratively using $V^{(n)}, \pi^{(n)}$ and $y^{(n)}$ for $(n-1)\rightarrow n$ with the following \textbf{actor-critic} scheme:
        \begin{enumerate}
            \item \textbf{Actor}: given $V^{(n-1)}$ we wish to find an optimal neural network
            policy~$\pi^{(n)}:\calX\times \calM\rightarrow \R^n$ which maximizes for all states $\ms=(z,\mm)\in\calS$ the expression
    \begin{equation}\label{eq:dhb_it_1}
    (TV^{(n-1)})(z,\mm) := \sup_{\pi}:\  U\!\left[  \,   \beta(\mm)\, V^{(n-1)}\big(\, 
               z'  + \pi(z,\mm)\cdot \mh';\, \mM '\,
                    \big)+ 
            R(\pi;z,\mm,\mM')\,
                    \Big|\,\mm\,
            \right]
            \ .
    \end{equation}
    We recall that $c(\ma;z,\mm)=\infty$
    whenever $\ma\not\in\calA(z,\mm)$. Hence, any finite solution $\pi^{(n)}$
    will be a member of~$\calP$.
    
        In the case of our OCE
        monetary utility, we will need to find both a network~$\pi^{(n)}$
        and a network~$y^{(n)}$ to jointly maximize for all states~$\ms=(z,\mm)$ i.e:
    \begin{equation}
        \sup_{\pi,y}: \  \E\!\left[\,\beta (\mm)\,
            u\!\left(  \,    V^{(n-1)}(\, z'  + \pi(z,\mm)\cdot \mh';\, \mM'\, ) + y(z,\mm) 
            \right) - y(z,\mm)        +  R(\pi;z,\mm,\mM')
                    \,\Big|\,\mm\,
            \right] \ .
    \end{equation}
    Following~\cite{SUTTON} will approach this by stipulating
    that we have a density~$\Q$ over
    all samples from~$\calS=\calX\times\calM$. In practise, $\calS$ 
    is a finite sample set, and $\Q$ could be
    a uniform distribution. We note that under $\Q$ the current portfolio
    and the current market are random variables $Z$ and $\mM$, respectively.
    The associated unconditional expectation is denoted by
    $\E[ \cdot ] = \int \Q[d\ms]\, \E[\cdot|\ms]$. 
    
    We then solve
    \begin{equation}\label{eq:dhb_it_1n}
        \sup_{\pi,y}: \ \E\!\left[\
             u\!\left(  \,  \beta (\mM)\  V^{(n-1)}(\, Z + \pi(Z,\mM)\cdot \mH';\, \mM'\, ) + y(Z,\mM) 
            \right) 
            - y(Z,\mM)  + 
            R(\pi;Z,\mM,\mM')\
            \right]  \ .
    \end{equation}
    Under~$\Q$
     the current market state, the portfolio and the hedging
     instrument representation
    are random variables, hence we have referred to them
    with capital letters. 

    The choice of~$\Q$ is not trivial: it represents the 
    probability of possible portfolio and market states.

    \item \textbf{Critic (Interpolation)}: as next step, we estimate a new value function~$V^{(n)}$ 
            given~$\pi^{(n)}$ and~$y^{(n)}$.
            This means fitting a neural network $V^{(n)}$ such that     
    \begin{equation}\label{eq:dhb_it_2}
 		V^{(n)}( z,\mm ) \equiv ( TV^{(n-1})( z, \mm )
 	\end{equation}
    We note that solving~\eqref{dhb_it_1n} numerically
    with packages like TensorFlow or PyTorch will also yield
    samples $TV^{(n-1)}(z,\mm)$ for all $(z,\mm)\in\calS$. Assuming this is the case we
    may find network weights for~$V^{(n)}$ by solving an interpolation problem of the form
  	\[
 		\inf_V: \E\!\left[
 		\
 			d\left( - V( Z,\mM ) + (TV^{(n-1)})( Z, \mM ) 
 			\right)\
 		\right]
 	\]  
 	for some distance $d$
 	over our discrete sample space. An example is $d(\cdot)=|\cdot|$.
 	Instead of using neural networks for the last step we may also
 	consider classic interpolation techniques such as kernel interpolators.
 	
        \end{enumerate}

        This scheme is reasonably
         intuitive as it iteratively improves the
        estimation of the monetary utility $V^{(n)}$ and the optimal
        action $a^{(n)}$. There is a question on how many training epochs
        to use when solving, in each step, for the action and the value function.
        In~\cite{SUTTON} there is a suggestion that using just \emph{one} step is sufficient.
        The authors call this the \textbf{actor-critic} method. There are several
        other discussions on the viability of such methods, see also~\cite{A3C}
        and the references therein.

		\begin{remark}
			In some applications we may not be able to 
			use samples of $TV^{(n-1)}$ to solve~\eqref{dhb_it_2},
			but make use of trained $a^{(n)}$ and $y^{(n)}$ directly.
						We therefore may solve
			   \begin{equation}\label{eq:dhb_it_2n}
       \inf_V: \E\left[ \left(
            - V(\,Z;\,\mM\,) +
        \E\! \!\left[\,  \beta (\mM)\ u\!\left(
                    V^{(n-1)}(\,\cdots; \mM'\, ) + y^{(n)}(Z,\mM) \right) - y^{(n)}(Z,\mM)
            + 
            R(\pi^{(n)};Z,\mM,\mM')
                    \,\Big|\,Z,\mM\,
                \right] 
            \right)^2 \right] \ .
    \end{equation}
        The nested expectation is numerically sub-optimal. In order
        to address this, we solve instead the unconditional
    \begin{equation}\label{eq:dhb_it_2n2}
       \inf_V: \E\left[ \left(
            - V(\,Z;\,\mM\,) +
            \beta (\mM)\ u\!\left(
                    V^{(n-1)}(\, \cdots, \mM'\, ) + y^{(n)}(Z,\mM) \right) - y^{(n)}(Z,\mM)
            + 
            R(\pi^{(n)};Z,\mM,\mM')
            \right)^2 \right) \ .
    \end{equation}
    which has the same gradient in~$V$, and therefore the same optimal solution.\footnote{
        \begin{proof}
        Assume that $V(\ms) \equiv V(\theta;\ms)$ where $\theta$ are our network parameters.
        Denote by~$\partial_i$ the derivative with respect to the $i$th parameter.
        Our equation then has the form $\inf_\theta f(\theta)$ where
        \[
            f(\theta) := \E\!\left[ ( V(\theta;\mS) 
                    + \E[ h(\mS') | \mS ] +g(\mS) )^2 \right]
        \]
        The gradient is
        \[
            \partial_{\theta_i} f'(\theta) = 2\, \E\!\left[ \partial_i V(\theta;\mS) ( V(\theta;\mS) + \E[ h(\mS') | \mS ] 
                    + g(\mS) ) \right] 
                    =
           2\, \E\!\left[ \partial_i V(\theta;\mS) ( V(\theta;\mS) + h(\mS') 
                    + g(\mS) ) \right] 
        \]
        Therefore $f$ has the same gradient as $
             \theta \mapsto
            \E\!\left[ ( V(\theta;\mS) + h(\mS') 
                    + g(\mS) )^2 \right] 
        $.
        \end{proof}}
		\end{remark}

    \subsection{Representing Portfolios}

        The most obvious challenge when applying the approach presented in section~\ref{sec:iterativeDBH}
        is the need to represent our portfolio is some numerically efficient way.
        The following is an extension of the patent~\cite{PATENT_DH}
        where we proposed using a more cumbersome signature representation of our trader instruments
        a'la~\cite{Lyons2019NonparametricPA}. 
        
        Assume that we are given historic market data $\mm_t$ 
        at time points~$\tau_0,\ldots,\tau_N$.
        Further assume that at each point $\tau_j$ we had in our book instruments
        $\mx^t=(x^{t,1},\ldots, x^{t,m_t})$ with $x^{t,i}\in\calX$.
        
        As~$\mx$ were actual historic instruments,
        we have for each $x^{t,i}$ 
        a vector $\mf^{t,i}_t \in \R^F$ 
     	of historic risk metrics computed in~$t$, such as the 
         book value, a range of greeks, scenarios and other 
         calculations made in~$\tau_t$ to assist humans in their risk 		management decisions.
       We assume that those metrics        $\mf^t_t=(\mf^{t,1}_t,\ldots,
        \mf^{t,m_t}_t)$
are also available for \emph{the same instruments}
       at the next time step~$\tau_{t+1}$, denoted by~$\mf^t_{t+1}$.
       Instrument which expire between $\tau_t$ and $\tau_{t+1}$
       will have their book value and
       all greeks and scenario values set to zero.
       
       It is a reasonable
        assumption that those metrics~$\mf$  have decent predictive power for 
        the behaviour of our instruments; after all this is what
        human traders use to do drive their risk management decisions. Hence we will use them as \textbf{instrument
        features}. We will here only consider linear features such that
        for any weight vector $\mw \in \R^{m_t}$ the feature vector (the greeks, scenarios etc)
        of the weighted instrument $\mw \cdot \mx^t$ is correctly given
        as $\mw\cdot \mf^t$, so there is no need to recompute it
        later.\footnote{We note that this linearity is satisfied for all common risk metric calculations
        except VaR and counterparty credit calculations.} 
        We have referred to such a representation in~\cite{PATENT_DH} as
        \emph{Finite Markov Representation}, or short \emph{FMR}. 
        
        We further denote by $r^i_t$
        the historic aggregated cashflows of $x^{t,i}$
        over the period~$[\tau_t,\tau_{t+1})$, all in our accounting currency.        
        We set $\mr_t := (r^1_t,\ldots,r^{m_t}_t)$. The aggregated cashflows
        of a weighted instrument $\mw \cdot \mx^{(t)}$ are~$\mw\cdot \mr_t$.
        Similarly, we use $\mB^t_u=(\mB^{t:1}_u,\ldots,\mB^{i,m_t}_u)$
        to refer to the book values of our instruments in $u\in\{t,t+1\}$,
        respectively.

        We also assume that we have for all our hedging instruments  access to
        their respective feature vectors $\mf^{h:t}_t$ for both~$\tau_t$
        and~$\tau_{t+1}$. 
        It is important to recall that the greeks $\mf^{h:t,i}_{t+1}$
        refer to the features of the $i$th hedging instrument traded
        at~$\tau_t$, but computed at~$\tau_{t+1}$. That means in particular
        $\mf^{h:t,i}_{t+1}\not=\mf^{h:t+1,i}_{t+1}$
        as the instrument mechanics changes between time steps.
        We also denote by $\mb^{h:t}_u$ the book values of our hedging instruments
        for $u\in\{t,t+1\}$.
        
        In addition to our instrument features, we also assume that we chose a
        reasonable subset of \textbf{market features} at each time step~$\tau_t$.
        We continue to use the symbol~$\mm$ for those features even though in practise
        we will not use the entire available state vector.\\

        \noindent
        We will now generate random scenarios as follows
        \begin{enumerate}
            \item Randomly choose $t\in\{0,\ldots,N-1\}$, which determines
            the market states~$\mm:=\mm_t$
            and $\mm':=\mm_{t+1}$.
            
            \item
            Identify the hedging instruments $\mh$ with their 
            finite Markov representation
            \[
            \begin{array}{llcl}
                 \mbox{Terminal FMR of hedging instruments}  & \mh'       & :=&  \mf^{h:t}_{t+1} \\
                \mbox{Book values for our hedging instruments} & \mB(\mh,\mm) & :=&  \mb^{h:t}_t \\
                   & \mB(\mh,\mm') & :=&  \mb^{h:t}_{t+1} \\
                \mbox{Cashflows of our hedging instruments} & \mr(\mh,\mm) & :=&  r^{h:t}_t \\
                \mbox{Cost}       & c(\ma;z,\mm) & \leftarrow & \ms_t, \mf^h_t
            \end{array}
            \]
            The concrete implementation of the last line depends on the specifics of
            the cost function. For example, proportional transaction cost on net 
            traded feature exposure
            are implemented using a weight vector~$\mgamma\in\R^F$
            and setting~$c(\ma;z,\mm) := |\ma \cdot (\mgamma\,\mf^t_t) |$.
            
            \item Choose a random weight vector~$\mw\in\R^{m_t}$
            and define a sample portfolio as $z:=\mw \cdot \mx$ with
            \[
            \begin{array}{llcl}
                \mbox{Initial and terminal FMR of the portfolio} & z    & :=&  \mw\cdot \mf^t_t \\
                                                                 & z' & :=&  \mw\cdot \mf^t_{t+1} \\
                \mbox{Book value of our portfolio} & B(z,\mm) & :=&   \mw\cdot \mb^t_t\\
                   & B(z,\mm') & :=&  \mw\cdot \mb^t_{t+1} \\
                \mbox{Cashflows of the portfolio} & r(z,\mm) & :=&  \mw\cdot \mx_t \ .
            \end{array}
            \]
            
            The construction of a reasonable randomization of the weight
            vector is important: if the samples are too different from likely
            portfolios, then the resulting model will underperform. 
            However, if only historic portfolios are used, then the model
            is less able to learn handling deviations. More importantly, though,
            generating portfolios increases sample size.

        \end{enumerate}
        This approach allows us training our actor-critic model with real data scenarios without the 
        need of a market simulator. Indeed, the use of a market simulator is difficult for this
        particular set up as it would require computing book values and greeks for simulated data for a large
        number of made up derivative instruments. An open research topic is whether it is feasible to write
        a simulator for the data sets generated above, e.g.~synthetically generating returns
        of market data jointly with the feature vectors of instruments.

	\section{Relation to Vanilla Deep Hedging}
		
		We will now discuss the relation of 
		equation~\eqref{bell_U} 
		 to the solution of a corresponding vanilla Deep Hedging
		Problem. 
			
		We start by stating our original Deep Hedging problem~\cite{DH}
		adapting the notation used here so far.
		We fix some initial time~$t=0$
		 with market state~$\mm\equiv \mm_0 \equiv \mM_0$.
		Subsequent market states are denoted by $\mM_{t+1} := \mM_t'$.
		We use a similar notation for all other variables. We use $\calT(\mM_t) := t$
		as the operator which extracts from a state $\mM_t$ current calendar time.
		We also define  the stochastic discount factor to zero 
		as~$\beta_t:=\beta(\mM_{t-1}) \beta_{t-1}$ starting with~$\beta_0:=1$.
		We note that $\beta_1 = \beta(\mm)$ and $\beta_{\calT(\mM_t)} = \beta(\mM_{t-1}) \beta_{\calT(\mM_{t-1})}$.
		
        For this part we will need to assume that 
		every hedging instrument has a time-to-maturity less than~$\tau^*$ in the
		sense that if we by $\ma\cdot \mh^t$ at time~$t$, then all the book value and all cashflows
		from the portfolio beyond $t+\tau^*$  are zero. This assumption excludes perpetual assets
		such as shares or currencies. We will need to trade those with futures or
		forwards in the current setup.

        Assume then that we are starting with an initial portfolio $z$ and follow a trading
        policy $\pi$. Accordingly, $Z_0 := z$ and $Z_{t+1} := Z_t' +\mA_t \cdot \mh^t{}'$
        where $\mA_t := \pi(Z_t,\mM_t)$. We also use $\mS_t:=(Z_t,\mM_t)$ where convenient. Assume the portfolio has 
        maturity~$T^*$ beyond which all cashflows are zero.
        
        The total gains from trading $\pi$ starting in $z$ are given as
        \begin{eqnarray}
            G^\pi(z)&  := & \sum_{t=0}^\infty \beta_t\label{eq:discRDH}
						R(\mA_t;Z_t,\mM_t,\mM_{t+1} )\\
						& = &  	\nonumber
						\underbrace{
						- \mB(z,\ms_0)
+
                            \sum_{t=0}^{T^*}
						        \beta_t r(z,\mM_t)
						  }_{\mbox{P\&L from $z$}}    
\\
&&						 + \nonumber
						  \sum_{t=0}^{\infty}
						    \beta_t
						    \Big( \underbrace{
						        - \mA_t\cdot \mB( \mh^t, \mM_t )
						        - c(\mA_t,\mS_t)
						         }
						    _{\begin{array}{c}
						        \mbox{Cost of trading}\\
						        \mbox{$\mA_t\cdot \mh^t$ in~$t$}
						        \end{array}}
						        +
						        \underbrace{
						        \mA_t\cdot 
						        \sum_{u=t}^{t+\tau^*}
						            \frac{\beta_u}{\beta_t}
						                r(\mh^t,\mM_u)
						        }_{\begin{array}{c}
						        \mbox{All future rewards}\\
						        \mbox{from trading $\mA_t\cdot \mh^t$ in~$t$}
						        \end{array}}
						    \Big)
        \end{eqnarray}
        We now introduce the  discounted cashflow rewards
       \[
       \hat R(\ma;z,\mm) := \beta_{\calT(\mm)}\, \tilde R(\ma;z,\mm)  \supstack{\eqref{discrewards}}=  \beta_{\calT(\mm)}\, \left(\ 
        r(z+\ma \cdot \mh,\mm) - \ma \cdot B(\mh,\mm) - c(\ma,z,\mm)
       \ \right)
       \]
        such that
        \[
            G^\pi(z) = \sum_{t=0}^\infty \hat R(\mA_t;Z_t,\mM_t)  \ .
        \]
        We say that the market has \emph{only finite statistical arbitrage} if we cannot make an infinite amount of money by trading
        from an empty portfolio in our hedging instruments. Formally,
        \[
            \infty > \gamma := \sup_\pi \E\!\left[\, G^\pi(0) \,\right]  \ .
        \]
        \begin{prop}
        If the market has only finite statistical arbitrage, then $U[\,G^\pi(z)\,]\leq \E[\,G^0(z)\,] + \gamma<\infty$ for all integrable $z\in\calX$ and
        all trading policies~$\pi\in\calP$.\footnote{
            Since $U$ is risk-averse we have
        $U[G^\pi(z)]=U[\,\sum_{t=0}^\infty \beta_t
						R(\mA_t,\mS_t ) ] \leq \E[ \sum_{t=0}^\infty \beta_t
						R(\mA_t,\mS_t ) ] = \E[\,G^0(z)\,] + \gamma< \infty$.\qed
						}
        \end{prop}
        That means the following definition makes sense:

        \begin{definition}
		The value function of the \textbf{Vanilla Deep Hedging} problem for an infinite trading
		horizon expressed as in~\cite{DH} in units of the underlying numeraire is given as
		the finite
		\eq{dh}
			\hat U^*(z,\mm) := 
			\sup_\pi:\ 
				U\!\left[\
					\sum_{t=0}^\infty 
					\hat R(\pi;Z_t,\mM_t )
				\ \Big|\ Z_0=z,\mM_0=\mm\ \right]
		\eqend
		The actual cash value function is given as
		\eq{dh_val_function_cash}
		    U^*(z,\mm) := \hat U^*(z,\mm) \ /\ \beta_{\tau(\mm_0)}
		\eqend
        \end{definition}
        Simple arithmetic\footnote{
        We may write out
            \eqary
            U^*(z,\mm) & = & \sup_\pi 	:\ 
				U\!\left[\
					\sum_{t=0}^\infty 
					\hat R(\pi;Z_t,\mM_t )
				\ |\ Z_0=z,\mM_0=\mm\ \right]
            \\
            & = &
            \sup_{\ma\in\calA} 	:\ 
				U\!\left[U\!\left[\
					\sum_{t=1}^\infty 
					\hat R(\pi;Z_t,\mM_t )
				\ |\ Z_1,\mM_1\ \right] + \hat R(\ma,Z_0,\mM_0)\,|\,Z_0=z,\mM_0=\mm\,\right]
				\\
            & = &
            \sup_{\ma\in\calA} 	:\ 
				U\!\left[U^*( Z_1,\mM_1) + \hat R(\ma,Z_0,\mM_0)\,|\,Z_0=z,\mM_0=\mm\,\right]
        \eqaryend
        \qed
        } shows that the value function for Deep Hedging solves 
        a Bellman equation:

        \begin{theorem}[Vanilla Deep Hedging Bellman Equation]
        Assume that the market is strictly free of statistical arbitrage, and that
        $U$ is time-consistent (i.e.~it is the entropy or the expectation). Then the  value function  $\hat U^*$ relative
       to the underlying numeraire satisfies
        the discounted dynamic programming equation
    \begin{equation}\label{eq:bell_U_DH}
    \left\{
    	\begin{array}{lcl}
       \hat  U^*\big(\,
            z;\,\mm
        \,\big)
        & \supstack!= &
        (\hat T \hat U^*)(z,\mm) \\
        &&
        \\
        (\hat Tf)(z,\mm) & := &
        \sup_{\ma\in \calA(z,\mm)}:\
           U\!\left[\ 
                   f\big(\,
            z' +\ma\cdot \mh';\,\mM'
        \,\big) 
            \,\big|\,\mm
            \ \right] + \hat R(\ma;z,\mm) \ .
       \end{array}
       \right.
       \end{equation}
       for discounted rewards~$\hat R$.
       
       The cash value function $U^*(z,\mm):=\hat  U^*(z,\mm)/\beta_{\calT(\mm)}$ satisfies
    \begin{equation}\label{eq:bell_U_DH_df}
    \left\{
    	\begin{array}{lcl}
         U^*\big(\,
            z;\,\mm
        \,\big)
        & \supstack!= &
          (\tilde T U^*)(z,\mm) \\
        &&
        \\
         (\tilde Tf)(z,\mm) & := &
        \sup_{\ma\in \calA(z,\mm)}:\
          \frac1{\beta_{\calT(\mm)}}\  U\!\left[\ \beta_{\calT(\mM')}
                   f\big(\,
            z' +\ma\cdot \mh';\,\mM'
        \,\big) 
            \,\big|\,\mm
            \ \right] + \tilde R(\ma;z,\mm) \ .
       \end{array}
       \right.
       \end{equation}
       (note the presence of the non-discounted rewards $\tilde R$ instead of $\hat R$).
       \end{theorem}
       
       The first observation we make is that~\eqref{bell_U_DH_df} only reduces to our original~\eqref{bell_U} 
       if~$U$ is coherent. In this case, we can move $\frac1{\beta_{\calT(\mm)}}$ inside~$U$.
       However, we have already seen that the value function of the vanilla Deep Hedging problem only
       solves~\eqref{bell_U_DH_df} if~$U$ is the entropy or the expectation. Since the entropy
       is not coherent, this means
       \begin{corollary}
        The cash value function~$U^*$ is only a solution to the Deep Bellman Hedging problem~\eqref{bell_U}  if $U=\E$.
       \end{corollary}
       
       \subsection{Solutions to the Vanilla Deep Hedging Bellman Equation}
       
       As in theorem~\ref{th:convergence}, we now reverse the situation and ask under which circumstances~\eqref{bell_U_DH_df}
       has a finite solution: recall that $\beta_{\calT(\mM_t)} = \beta(\mM_{t-1}) \beta_{\calT(\mM_{t-1})}$
       
       \begin{theorem}[Existence of finite unique Solutions for the Vanilla Deep Hedging Bellman Equation]\label{th:convergence_dh}
        Assume that rewards $\tilde R$ are finite and that $U$ is a monetary utility.
        
        Then, the Vanilla Deep Hedging Bellman Equation~\eqref{bell_U_DH_df} has a unique finite solution.
        
        If $U$ is the entropy or the expectation, then this optimal solution coincides with the
        cash value function~\eqref{dh_val_function_cash} of the Vanilla Deep Hedging problem.
       \end{theorem}
       The proof for the existence of a unique finite solution is presented in section~\ref{sec:proof_dh}.
       
       In light of this result it is evident that both the Deep Bellman Hedging equation~\eqref{bell_U} and
       the Vanilla Deep Hedging Bellman equation~\eqref{bell_U_DH_df}
       are reasonable candidates for solving the generic hedging problem. However, the latter requires
       us to essentially fix an initial point point $t=0$, upon which the numeraire $\beta_t$ is based.
       In order to maintain consistence accross time, that initial time point would need to be kept 
       constant and therefore in the distant past.
       We therefore recommend
       using~\eqref{bell_U} as presented.

    \section{Existence of a  Unique Finite Solution for Deep Bellman Hedging}\label{sec:proofs}\label{sec:proof_convergence}

		We will now prove with theorem~\ref{th:convergence} 
		convergence of our Deep Bellman Hedging
		equations. This is easiest understood when the space~$\calZ$ of 
		future cashflows is parameterized in $\R^{|\calZ|}$
		with a finite Markov representation. However, in more generality
		we may assume that $\calZ$ represents the set of suitably
		integrable 
		adapted stochastic processes with values in~$\R$.
		Therefore, we may 
		just assume that $(\calS,\Q)$ with $\calS=\calZ\times\calM$ 
		is a measure space. In the following we will
		consider the function space $F$ of the $\Q$-equivalence
		classes of functions $f:\calS\rightarrow \R$.
        
         Let as before
        \eq{tf_proof}
            (Tf)(z,\mm) := \sup_{\ma\in\calA(z,\mm)}:  \ 
            U\!\left[ \beta(\mm)  f\big( 
                z' +\ma\cdot\mh', \mM' \big) + R(\ma,z,\mm,\mM') \
                \big|\mm\right]
        \eqend
        for $\beta(\mm)\leq\beta^*<1$.
        Then the Bellman equation $f=Tf$ has a unique, finite
         solution.

        We will demonstrate the proof for bounded value 
        functions. See~\cite{UNBOUNDED} on how 
        to extend results of convergence of Bellman operators 
        to the unbounded case. The below mimics the spirit
        of the proof of the classic Banach contraction theorem.\\
        
        \begin{proof}[of theorem~\ref{th:convergence}]
        equip $F$ with the supremum norm. We wish to
        show that for $\|f\|_\infty<\infty$ we have $\| Tf \|<\infty$.
        
        Monotonicity and cash-invarianc yield
        \[
        	Tf \leq T \| f \| = T0 + \| f \| \ .
        \]
        Using~\eqref{rewards_finite} we therefore find that~$T0<\infty$.
        
        We will now show that~$Tf$ is a \emph{contraction}
        for bounded~$f:\calS\rightarrow \R$,
         i.e.~$\| Tf - Tg \|\leq \beta^* \| f - g \|$
  		for our $\beta^*<1$. Note that $f(\ms)-g(\ms)\leq \| f - g \|$.
        Monotonicity and cash invariance of the operator~$T$ yield
        \[
            (Tf)(\ms) \leq T\left( g + \| f - g \| \right)(x) \leq (Tg)(\ms) + \beta^*\| f-g \| 
        \]
        Similarly,
        \[
            (Tg)(\ms) \leq T\left( f + \| f - g \| \right)(x) \leq (Tf)(\ms) + \beta^*\| f-g \| \ .
        \]
        Jointly this gives
        \[
            \| Tf - Tg \| \leq \beta^*\| f-g \| \ .
        \]
        Applying the Banach fixed-point theorem then yields the desired result.\footnote{ 
        For illustration, we sketch a simple proof:
        Chose $f_0$ and 
        let $f_n := Tf_{n-1}$
        such that $f_n = T^n f_0$.
        We know that $\| Tf_1 - Tf_0 \|
            \leq \beta^* \| f_1 - f_0 \|
        $
        and therefore iteratively
        $\| Tf_n - Tf_{n-1} \|
            \leq \beta^*{}^n \| f_n - f_{n-1} \|
        $.
        Triangle inequality implies $\| Tf_n - Tf_m \|
        \leq \sum_{i=m+1}^n
        \| Tf_i - Tf_{i-1} \|
        \leq \| f_1 - f_0 \| \sum_{i=m+1}^n \beta^*{}^i
        \downarrow 0$. This means $Tf_n$ is a Cauchy
        sequence and therefore converges to a unique point
        $f_n\rightarrow f$.
        
        To show that $f$ is a fixed point
        note that $\| Tf - f | \leq \| Tf - f_n \|
        + \| f_n - f \| 
        \leq \beta^* \| f - f_{n-1} \|  + \| f_n - f \|  \downarrow 0$.
        }
        \end{proof} 
        
        \subsection{Vanilla Deep Hedging Bellman Equation}\label{sec:proof_dh}

        We now focus on \eqref{bell_U_DH_df} i.e.~the operator
       \[
       (\tilde Tf)(z,\mm)  :=
        \sup_{\ma\in \calA(z,\mm)}:\
           \frac1{\beta_{\calT(\mm)}}\  U\!\left[\ \beta_{\calT(\mM')}
                   f\big(\,
            z' +\ma\cdot \mh';\,\mM'
        \,\big) 
            \,\big|\,\mm
            \ \right] + \tilde R(\ma;z,\mm) \ .
       \]
        \begin{proof}[of theorem~\ref{th:convergence_dh}]
       With our previous assumptions is clear that $\tilde T0 < \infty$. 
       We now show that $\tilde T$ admits to a discounted form of cash invariance in the sense that
       \[
          T (f(\cdot)+y(z,\mm))(z,\mm) = (Tf)(z,\mm) + \beta(\mm) y(z,\mm)
       \]
       To this end, recall that $\beta_{\calT(\mM')} = \beta(\mm) \beta_{\calT(\mm)}$. Hence,
       \eqary
        \left(\,T(\, f(\cdots) + y(z,\mm) \,)\,\right)(z,\mm)
        & =&  
         \sup_{\ma\in \calA(z,\mm)}:\
           \frac1{\beta_{\calT(\mm)}}\  U\!\left[\ \beta(\mm) \beta_{\calT(\mm)}
                  \left( f(\cdots) + y(z,\mm) \right)
            \,\big|\,\mm
            \ \right] + \tilde R(\ma;z,\mm) 
        \\
        & \supstack{(*)}= &
         (Tf)(z,\mm) + \beta(\mm) y(z,\mm) \ .
       \eqaryend
       The last equality follows from cash-invariance of the operator~$U$.
       
       Hence, just as in the preceding proof,
        \[
            (\tilde Tf)(\ms) \leq \tilde T\left( g + \| f - g \| \right)(x) \leq (\tilde Tg)(\ms) + \beta^*\| f-g \| 
        \]
        which shows that $\tilde T$ is also a contraction. Applying the Banach fixed-point theorem
        yields the existence of a unique solution.\end{proof}

	\subsection*{Disclaimer}
	
	Opinions and estimates constitute our judgement as of the date of this Material, are for informational
purposes only and are subject to change without notice. It is not a research report and is not intended
as such. Past performance is not indicative of future results. This Material is not the product of J.P.
Morgan’s Research Department and therefore, has not been prepared in accordance with legal requirements
to promote the independence of research, including but not limited to, the prohibition on the dealing ahead
of the dissemination of investment research. This Material is not intended as research, a recommendation,
advice, offer or solicitation for the purchase or sale of any financial product or service, or to be used in
any way for evaluating the merits of participating in any transaction. Please consult your own advisors
regarding legal, tax, accounting or any other aspects including suitability implications for your particular
circumstances. J.P. Morgan disclaims any responsibility or liability whatsoever for the quality, accuracy or
completeness of the information herein, and for any reliance on, or use of this material in any way.
Important disclosures at: www.jpmorgan.com/disclosure

    % -----------------------------------------------------------------------------------
    % -----------------------------------------------------------------------------------

   \addcontentsline{toc}{section}{Bibliography}

    \bibliographystyle{alpha}
    \bibliography{references}

\end{document}